\documentclass[sigplan,nonacm,twocolumn]{acmart}
\renewcommand\footnotetextcopyrightpermission[1]{}
\usepackage{enumitem}
\usepackage{multirow}
\usepackage{tcolorbox}
\usepackage{cleveref}
\usepackage{mdframed}
\usepackage{caption}
\usepackage[compact]{titlesec}

\usepackage{xspace}
\usepackage{soul}
\usepackage{algorithm}
\usepackage[noend]{algpseudocode}

\newcommand{\mypar}[1]{\vspace{1ex}\noindent\textbf{\textit{#1}}}
\newcommand{\myparnospace}[1]{\vspace{.5ex}\noindent\textbf{\textit{#1}}}

\AtBeginDocument{%
  }

\acmSubmissionID{\#611}

\newcommand{\sysname}{ARMS\xspace}

\begin{document}


\title{\sysname: Adaptive and Robust Memory Tiering System}


  \author{Sujay Yadalam}
  \authornote{Both authors contributed equally to this work.}
  \email{sujayyadalam@cs.wisc.edu}
  \affiliation{%
    \institution{University of Wisconsin-Madison}
    \city{Madison}
    \state{WI}
    \country{USA}
  }

  \author{Konstantinos Kanellis}
  \authornotemark[1]
  \email{kkanellis@cs.wisc.edu}
  \affiliation{%
    \institution{University of Wisconsin-Madison}
    \city{Madison}
    \state{WI}
    \country{USA}
  }

  \author{Michael Swift}
  \email{swift@cs.wisc.edu}
  \affiliation{%
    \institution{University of Wisconsin-Madison}
    \city{Madison}
    \state{WI}
    \country{USA}
  }

  \author{Shivaram Venkataraman}
  \email{shivaram@cs.wisc.edu}
  \affiliation{%
    \institution{University of Wisconsin-Madison}
    \city{Madison}
    \state{WI}
    \country{USA}
  }

\begin{abstract}
Memory tiering systems seek cost-effective memory scaling by adding multiple tiers of memory. For maximum performance, frequently accessed (hot) data must be placed close to the host in faster tiers and infrequently accessed (cold) data can be placed in farther slower memory tiers. Existing tiering solutions such as HeMem, Memtis, and TPP use rigid policies with pre-configured thresholds to make data placement and migration decisions. These thresholds make the systems brittle - they fail to perform well in all scenarios. Our analysis of existing systems revealed that incorrect tiering parameters lead to: inaccurate hot page identification, delayed response to hot set changes, and wasteful migrations.

Based on this study, we designed \sysname that replaces sensitive parameters with robust policies and mechanisms. 
We develop a novel hot/cold page identification mechanism that uses relative scoring rather than threshold comparison, a hot set change detector to adapt to workload distribution changes, an adaptive migration policy based on cost/benefit analysis, and a bandwidth-aware batched migration scheduler. Combined, these approaches provide an out-of-the-box performance that matches the best tuned performance of prior systems, while being $1.22$-$1.85x$ better than prior systems without tuning.

\end{abstract}

\maketitle

\section{Introduction}
\label{sec:intro}

\begin{figure}
    \centering
    \includegraphics[width=\linewidth]{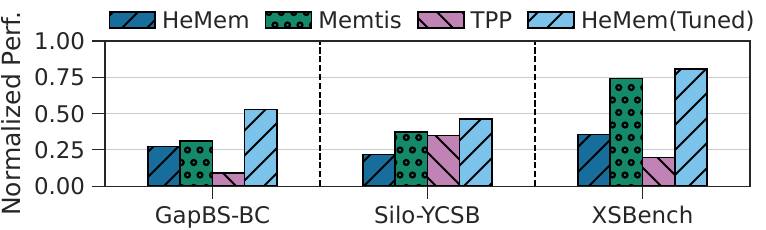}
    \caption{Out-of-the-box HeMem, Memtis, TPP performance vs \textit{tuned} HeMem performance. Performance normalized to all data in fast-tier (Higher is better).}
    \label{fig:intro_opportunity}
\end{figure}

Datacenters are increasingly adopting tiered memory systems to meet the memory demands of modern data-intensive applications such as graph processing, machine learning and in-memory databases. Tiered memory systems supplement existing DRAM with memory tiers made of new and/or cheaper memory technologies such as Non-Volatile Memory~\cite{nvm2018} or CXL-attached memory~\cite{cxlspec}. This enables systems with large amounts of memory to be built in a cost-effective manner. Application data is distributed across tiers and can be accessed directly from any tier without caching.

New memory tiers built using NVM or CXL have better price and/or density but worse performance than locally-attached DRAM. Therefore, repeatedly accessing data from these slow tiers can degrade application performance. Memory tiering systems avoid this by keeping frequently or recently accessed (hot) data in faster tiers, 
and the rest (cold) of the data in slow tiers.

Numerous works~\cite{hemem2021, memtis2023, flexmem2024, autotiering2017, thermostat2017, tmts2023, hybridtier2025, artmem2025, tpp2023, hmsdk2024} have been proposed to smartly place and migrate data in memory tiering systems, transparent to running applications. 
Unfortunately, we find that these solutions do not work well for all applications and system configurations. 
This is mainly because these systems use \textit{rigid} policies to make data placement decisions: they assume stable workloads, stationary signal scales, and fixed operating conditions. In practice, access measurements are noisy (PEBS sampling and PTE walks drop events), hot sets change over time, while application inputs and device characteristics also vary.
They inherit the rigidity due to their reliance on \textit{pre-configured thresholds} to make data placement decisions (e.g., promote if accesses $\ge \theta$ over a window; demote otherwise).
The values of such parameters are typically optimized for the common case and not tailored for specific workload behavior or configuration. 
As a result, we observe that these systems fail to
correctly identify all the hot pages and migrate them in time, and/or perform many unnecessary migrations.

Some systems such as Memtis~\cite{memtis2023} and ArtMem~\cite{artmem2025} attempt to address the above limitation by dynamically adjusting the \textit{hot\_threshold} ($\theta$) for page promotion to ensure that the correct set of hot pages is loaded in the fast tier. These systems still underperform because they use static thresholds for other important parameters such as cooling period and adaptation period. As shown in Table~\ref{table:relatedwork_parameters}, all existing systems use static parameters.

To demonstrate how much existing solutions suffer from the use of static knob values, 
we conduct a simple experimental study using HeMem~\cite{hemem2021}, Memtis~\cite{memtis2023} and TPP~\cite{tpp2023} on a machine equipped with NVM as the slow tier.
We first measure the performance of each system with their default, static knob values.
Then, we identify the best-performing set of knob values for HeMem, by tuning the values using Bayesian Optimization.
Figure~\ref{fig:intro_opportunity} shows the performance of HeMem, Memtis and TPP, normalized to the baseline of all data in DRAM.
While HeMem and Memtis provide speedups over TPP, the \textit{tuned} HeMem system yields significant improvements over them, showing that its use of static thresholds for all workloads significantly limits their performance.

While tuning is beneficial, it is expensive.
In this paper, we argue that for tiering systems to be robust, we need to avoid using fixed thresholds and adapt policies based on the workload and current system state.
To achieve this, we build a robust tiering system based on the following principles:
\begin{itemize}[topsep=0pt,leftmargin=1.5em]
    \item \textbf{Noise tolerance:} smooth input signals so that noise due to sampling errors or transient fluctuations does not lead to poor decisions.
    \item \textbf{Relative decision-making:} compare candidates against one another rather than against fixed thresholds.
    \item \textbf{State awareness:} account for the current system state, not just the workload, when making decisions.
    \item \textbf{Cost--benefit reasoning:} act only when a decision's estimated benefit outweighs its cost.
\end{itemize}
Based on the above, we build \textit{\sysname}, an \underline{A}daptive and \underline{R}obust \underline{M}emory tiering \underline{S}ystem that dynamically adjusts to both workload characteristics and the system state, consistently achieving high performance without the need for tuning.

First, \sysname improves the accuracy of hot/cold page identification through relative page scoring rather than absolute threshold comparison. To calculate page scores, \sysname uses access histories of different lengths: one short (recent hotness) and one long (historical hotness).
Maintaining two histories help \sysname balance between stability and responsiveness.
\sysname ranks application pages, and selects the top $k$ pages to be placed in the fast tier based on the scores, where $k$ is the fast tier capacity. 
Note that ranking is invariant to global level shifts and avoids workload-specific cutoffs. 

Second, \sysname uses a change point detection algorithm to identify changes in the workload access distribution. When changes occur, \sysname uses the short-term access average to quickly identify newly hot pages to promote to the fast tier.

Third, inspired by prior work~\cite{cbmm2022}, \sysname uses a cost-benefit model for migration decisions to help eliminate wasteful migrations. \sysname estimates the benefit of a page promotion using its access history (moving averages) and an estimation of how long it will be hot. It estimates the migration cost based on prior migrations and available bandwidth. A page is promoted only if the benefit exceeds the cost.

Last, \sysname makes two changes to the migration mechanism. It prioritizes promotion of the hottest pages over other hot pages and demotes the coldest pages eagerly. To speed up migrations, \sysname uses batched migrations, similar to Nimble~\cite{nimble2019}. \sysname dynamically adjusts the batch size based on the available bandwidth to ensure that migrations do not affect application throughput. Together, these four
optimizations remove the need for thresholds by dynamically responding to application behavior and system conditions.

We evaluate \sysname with a diverse set of workloads on different machines and configurations. On a machine with remote NUMA memory as the slow tier, we find that \sysname outperforms existing state-of-the-art tiering systems, including HeMem; Memtis; HMSDK; TPP; and HybridTier, by $1.22$-$1.85$x on average, without requiring tuning of any sort. Further, \sysname matches or slightly outperforms the best tuned configuration for HeMem, Memtis, and HMSDK.

In summary, we make the following contributions:
\begin{itemize}[topsep=0pt,leftmargin=0.7cm]
    \item We analyze how the use of static parameters affects tiering decisions and identify three key limitations of existing tiering systems.
    \item We identify design principles for robust tiering systems: smoothing inputs to reduce noise, adapting to current system conditions, calculating the expected value of actions, and making relative decisions based on this.
    \item We propose a new tiering system, \sysname, which embraces the above principles to make robust data placement decisions, while also adapting to the workload and system state, outperforming existing systems.
\end{itemize}
\vspace{-1ex}
\section{Background}
\label{sec:background}

\begin{table}[t]
\scriptsize
\begin{tabular}{@{}ll@{}}
\toprule
\multicolumn{1}{c}{\textbf{System}} & \multicolumn{1}{c}{\textbf{Static thresholds/parameters}} \\ \midrule
AMP~\cite{adaptivepagemigration2020}        & access\_history\_length, access\_update\_period  \\
Multi-Clock~\cite{multiclock2022}           & scan\_interval, active\_inactive\_list\_ratio    \\
AutoNUMA~\cite{autonuma}                    & scan\_interval, demotion\_watermark \\
HeMem~\cite{hemem2021}                      & hot\_threshold, cooling\_threshold               \\
TPP~\cite{tpp2023}                          & scan\_size, demotion\_watermarks                 \\
TMTS~\cite{tmts2023}                        & hot\_scan\_period, promotion\_threshold          \\
HMSDK~\cite{hmsdk2024}                      & nr\_regions, promotion\_accesses\_min            \\
Memtis~\cite{memtis2023}                    & adaptation\_interval, cooling\_interval          \\
FlexMem~\cite{flexmem2024}                  & adaptation\_interval, countdown\_timer           \\
HybridTier~\cite{hybridtier2025}            & momentum\_threshold, cooling\_period             \\ 
\bottomrule
\end{tabular}
\vspace{1ex}
\caption{Static thresholds/parameters used in existing tiering systems. We only list 2, yet most systems have at least 5.}
\label{table:relatedwork_parameters}
\vspace{-1ex}
\end{table}

\noindent
\textbf{\textit{Memory tiering.}} Modern applications such as in-memory databases, web serving, graph processing and machine learning demand large memory systems for high performance. 
Unfortunately, scaling memory naively has become challenging because: (1) DRAM scaling has stagnated over the past few years and the cost of DRAM has been increasing making memory costs a large portion of the Total Cost of Ownership (TCO)~\cite{tmo2022}, (2) the number of channels per socket is limited by the number of pins on the chip.
One way to achieve cost-effective memory scaling is through \textit{memory tiering}, which involves the organization of data across multiple \textit{tiers} of memory. Memory tiers are built using new memory technologies such as NVM~\cite{nvm2018} (lower cost/byte), CXL-based memories~\cite{cxlspec} or byte-addressable NVMe SSDs~\cite{xu2015performance}.

Researchers have proposed several approaches to smartly place data in memory tiers for achieving good performance~\cite{autotiering2017,thermostat2017,nimble2019,hemem2021,tpp2023,tmts2023,memtis2023,matryoshka2024,flexmem2024,colloid2024,kleio2019}. These memory tiering systems perform two key tasks: (1) \textit{identify hot and cold data/pages}, and (2) \textit{migrate the identified hot/cold pages between tiers}.

\mypar{Access monitoring.} Existing tiering systems rely on one or more of the following to identify hot and cold pages: 1) hardware event sampling such as Intel PEBS~\cite{pebs2018}, 2) page table scanning such as DAMON~\cite{damon2019} and/or 3) page fault tracking to identify hot and cold pages.
Hardware event sampling allows software to track hardware events such as LLC misses or processor writes. The hardware records the exact memory address related to the event. Sampling overhead is higher at higher sampling frequencies since more samples are collected and processed. 
DAMON scans the Accessed (A) bits in page table entries periodically to identify hot and cold pages. To overcome the overhead of scanning all page table entries (PTE), DAMON divides the address space into regions and monitors only sampled pages in each region.
Last, some tiering systems~\cite{thermostat2017,tpp2023,autonuma} use page faults (NUMA hint faults) to gather information about the set of pages touched by an application.
All these approaches are prone to inaccuracies, especially at lower sampling/scanning frequencies. Recent work~\cite{m5,neomem2024} address this by proposing new CXL-side hardware counters.

\mypar{Page hotness heuristics.} Existing systems use access frequency, recency, or a combination of both to classify pages as hot or cold; these heuristics directly drive migration.
Frequency-based systems place pages with higher access counts in the fast tier, typically over a sliding window or an exponentially decayed counter. They typically react slowly to shifts as they maintain historic access information and are sensitive to sampling rate and window length.
Recency-based systems adapt quickly to changes by placing recently accessed pages in the fast tier, but can misclassify pages with periodic or bursty access.
In practice, tiering implementations approximate these ideas with sampled counts, moving averages, and thresholding (\texttt{hot\_threshold}, \texttt{cooling\_threshold}, \texttt{sampling\_period}), often with simple hysteresis or minimum residency to curb thrash.
Because promotion/demotion policies are tightly coupled to these thresholds, inaccurate values can bloat the hot set, elongate promotion delay, and amplify oscillations under noise. 
\section{Motivation}

Existing tiering systems rely on weak signals of memory accesses (PEBS, page faults etc.) and use heuristics to make migration decisions.
We find that these systems perform poorly with some applications and system configurations due to one main reason: their heuristics use pre-defined static thresholds (Table~\ref{table:relatedwork_parameters}) to identify and migrate hot/cold pages across tiers. 
For example, HeMem relies on PEBS samples to detect whether a page is frequently referenced, but to keep overhead low it samples at a low frequency, which leads to inaccurate samples and random fluctuations. 

\subsection{Impact of parameter values on performance.}

\begin{figure}[t]
    \setlength{\abovecaptionskip}{1.5ex}
    \begin{minipage}{0.48\linewidth}
        \centering
        \includegraphics[width=\linewidth]{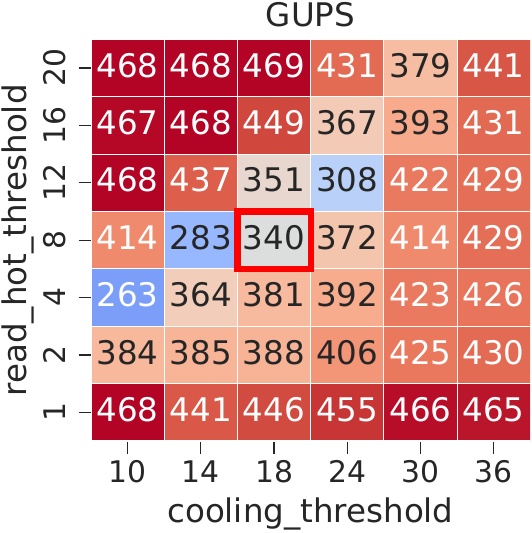}
    \end{minipage}
    \hfill
    \begin{minipage}{0.48\linewidth}
        \centering
        \includegraphics[width=\linewidth]{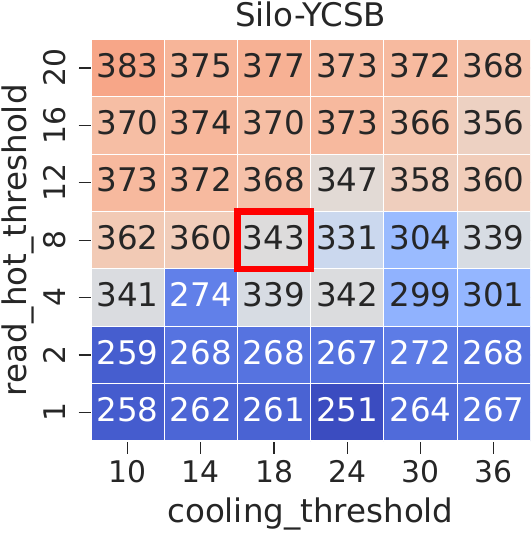}
    \end{minipage}
    \caption{Execution time (in seconds) of \textit{GUPS} (left) and \textit{Silo} (right) workloads, when we tweak two HeMem parameters. Default configuration execution time is shown in red box.}
    \label{fig:grid-search}
\end{figure}

To understand the extent to which thresholds affect memory tiering performance, we perform a simple study using HeMem~\cite{hemem2021} on a machine with Intel Optane NVM as the slow tier. We consider just two of HeMem's configuration parameters: \texttt{read\_hot\_threshold}, and \texttt{cooling\_threshold}. The \texttt{cooling\_threshold} determines when page access counts are halved to maintain freshness of the hot set estimation and adapt to changes in the working set. By default, HeMem sets this threshold to 18 sampled accesses.

Figure~\ref{fig:grid-search} shows the execution time for GUPS~\cite{plimpton2006simple} and Silo~\cite{tu2013speedy} (running a read-only YCSB-C workload) under different HeMem parameter configurations.
We highlight the default HeMem configuration with a red box.

From these results, we make the following observations.
First,  different thresholds result in significant variations in workload performance.
Second, there are some configurations that perform much better than the default HeMem configuration for a given workload.
For example, the best-performing configuration performs up to $29\%$ better than the default for GUPS and $36\%$ better for Silo.
Third, we observe that the ideal parameter values are quite different for GUPS (10, 4) and Silo (24, 1). In summary, we find that workloads are sensitive to pre-defined thresholds in memory tiering systems and tuning these parameters can provide significant performance benefits.

\begin{mdframed}[linecolor=black,linewidth=1pt,innerleftmargin=8pt,innerrightmargin=8pt,innertopmargin=4pt,innerbottommargin=4pt, skipabove=8pt]
\textbf{Insight:} \textit{Application performance is sensitive to parameters used in existing tiering systems. The best parameter values vary across applications.}
\end{mdframed}

\subsection{Parameter tuning study}
\label{sec:tuning_study}

To understand how and why certain parameter values lead to better performance than others, we perform a comprehensive study of tiering systems behavior with different workloads on different hardware.

To identify the best values quickly, we employ \textit{Bayesian Optimization} (BO), which has been shown to identify the best performing configurations in as few iterations as possible~\cite{snoek2012practical}.
We use the state-of-the-art Sequential Model-based Algorithm Configuration (SMAC) framework~\cite{hutter2011sequential}, previously used to tune other systems~\cite{kanellis2022llamatune}.
SMAC utilizes a Random Forest (RF) to model the parameter space; this enables it to handle large high-dimensional spaces efficiently.

We use a NUMA machine and treat remote NUMA memory as the slow tier. We run nine representative and diverse workloads (details in Section~\ref{sec:eval}). 

We tuned three systems: (1) HeMem~\cite{hemem2021}, which is a frequency-based policy with all static thresholds, (2) Memtis~\cite{memtis2023}, which uses frequency heuristic and dynamically adjusts the \texttt{hot\_threshold}, and (3) HMSDK~\cite{hmsdk2024}, which uses both access frequency and recency data from DAMON~\cite{damon2019} access monitoring.

\begin{figure}[t]
    \centering
    \includegraphics[width=0.95\linewidth]{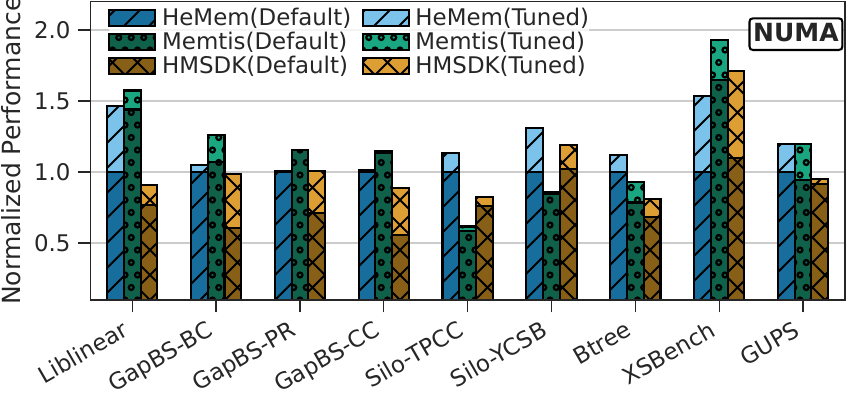}
    \caption{Performance of default vs tuned configuration for HeMem, Memtis and HMSDK. Normalized to default HeMem. Higher is better.}
    \label{fig:hemem-improv-scailp}
\end{figure}

\begin{figure}[t]
    \includegraphics[width=0.95\linewidth]{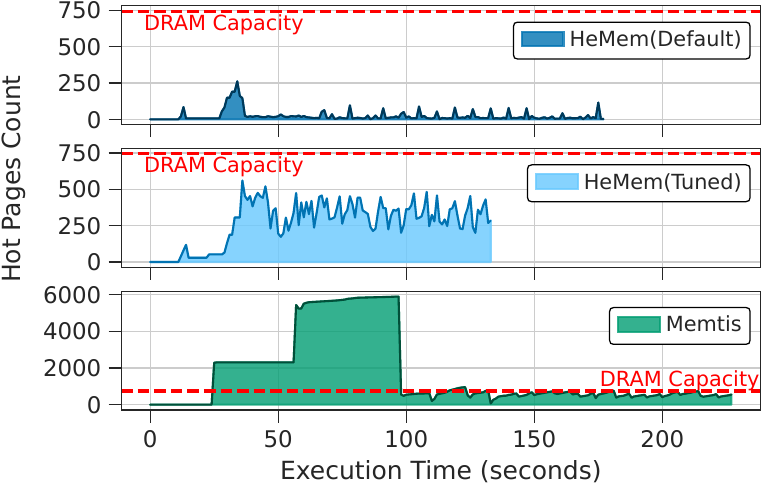}
    \vspace{.5ex}
    \caption{Number of pages classified as hot by HeMem and Memtis for \textit{GapBS-BC} over time. With default HeMem very few hot pages are identified. With tuned HeMem, most hot pages are identified. Memtis \textit{misclassifies} many pages as hot.}
    \label{fig:hotpageaccuracy}
    \vspace{-2ex}
\end{figure}

\mypar{Results and Analysis.}
\label{sec:tuning_analysis}
Figure~\ref{fig:hemem-improv-scailp} shows the performance benefit of parameter tuning. Tuning yields significant benefit to HeMem and HMSDK which uses all static thresholds. Memtis also benefits from tuning for some workloads. Overall, using the correct parameter values can yield $1.05-2.09x$ performance improvement.
By comparing the tiering decisions with the best parameters with that of the default parameters, we identify three symptoms of inaccurate parameter values:

\vspace{.25ex}
\textbf{1. Inaccurate hot page classification:}
If the \texttt{hot\_threshold} is not set correctly, HeMem and HMSDK fail to identify all the hot pages. Figure~\ref{fig:hotpageaccuracy} shows the pages classified as hot by HeMem with the default and best configurations for \textit{GapBS-BC}.
Similarly, an inaccurate \texttt{cooling\_threshold} causes HeMem to miss certain hot pages. If the value is too small, certain extremely hot pages frequently trigger page cooling. Due to such frequent cooling, the access counts of other (slightly less) hot pages do not reach the desired \texttt{hot\_threshold} value, and thus are never promoted.

\vspace{.25ex}
\textbf{2. Slow adaptation to changing hot sets:}
On the other hand, if the \texttt{cooling\_threshold} is set to high value, cooling is performed less frequently, which leads to a lot of pages being classified as hot. Figure~\ref{fig:hotpageaccuracy} shows how Memtis overestimates the set of hot pages due to an inaccurate \texttt{cooling\_threshold}.
Since no cold pages are identified in the fast tier, newly identified hot pages in the slow tier cannot be promoted.
As a result, these systems fail to adapt quickly to changing hot sets.

\begin{figure}[t]
    \centering
\includegraphics[width=.9\linewidth]{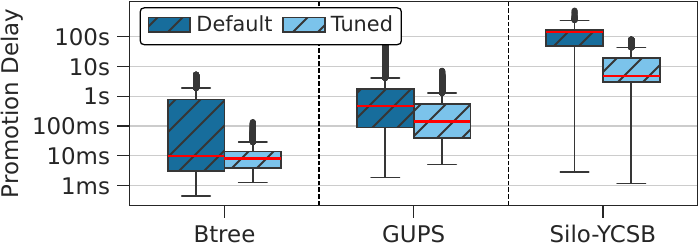}
    \caption{Promotion delay of hot pages for default and tuned HeMem configurations for \textit{Btree}, \textit{GUPS} and \textit{Silo-YCSB}. Box contains [25\%,75\%] of values. Red line indicates median delay.}
    \label{fig:promotion_delay}
\end{figure}

To showcase this issue, we measure the average \textit{promotion delay} of hot pages in HeMem, which corresponds to the time since a hot page is identified to when it is actually promoted. Figure~\ref{fig:promotion_delay} compares the promotion delay when using default and best configurations.
Interestingly, we observe that for workloads like GUPS and Silo, the average promotion delay can be extremely large, i.e., in the order of 100s of seconds.

These systems exacerbate the problem by performing migrations in a strictly serial manner, allowing only one page to migrate at a time.
Therefore, extremely hot pages added to the tail of the queue have to wait for all the previous queue entries to be promoted first, thereby significantly increasing their promotion delay.

\vspace{-1ex}
\begin{mdframed}[linecolor=black,linewidth=1pt,innerleftmargin=8pt,innerrightmargin=8pt,innertopmargin=4pt,innerbottommargin=4pt, skipabove=8pt]
\textbf{Insight:} \textit{Existing tiering systems struggle to balance accuracy and reactivity — static thresholds make them either accurate but sluggish, or responsive but noisy.}
\end{mdframed}
\vspace{1.5ex}

\begin{figure}[t]
    \centering
    \begin{minipage}{0.47\linewidth}
        \includegraphics[width=\linewidth]{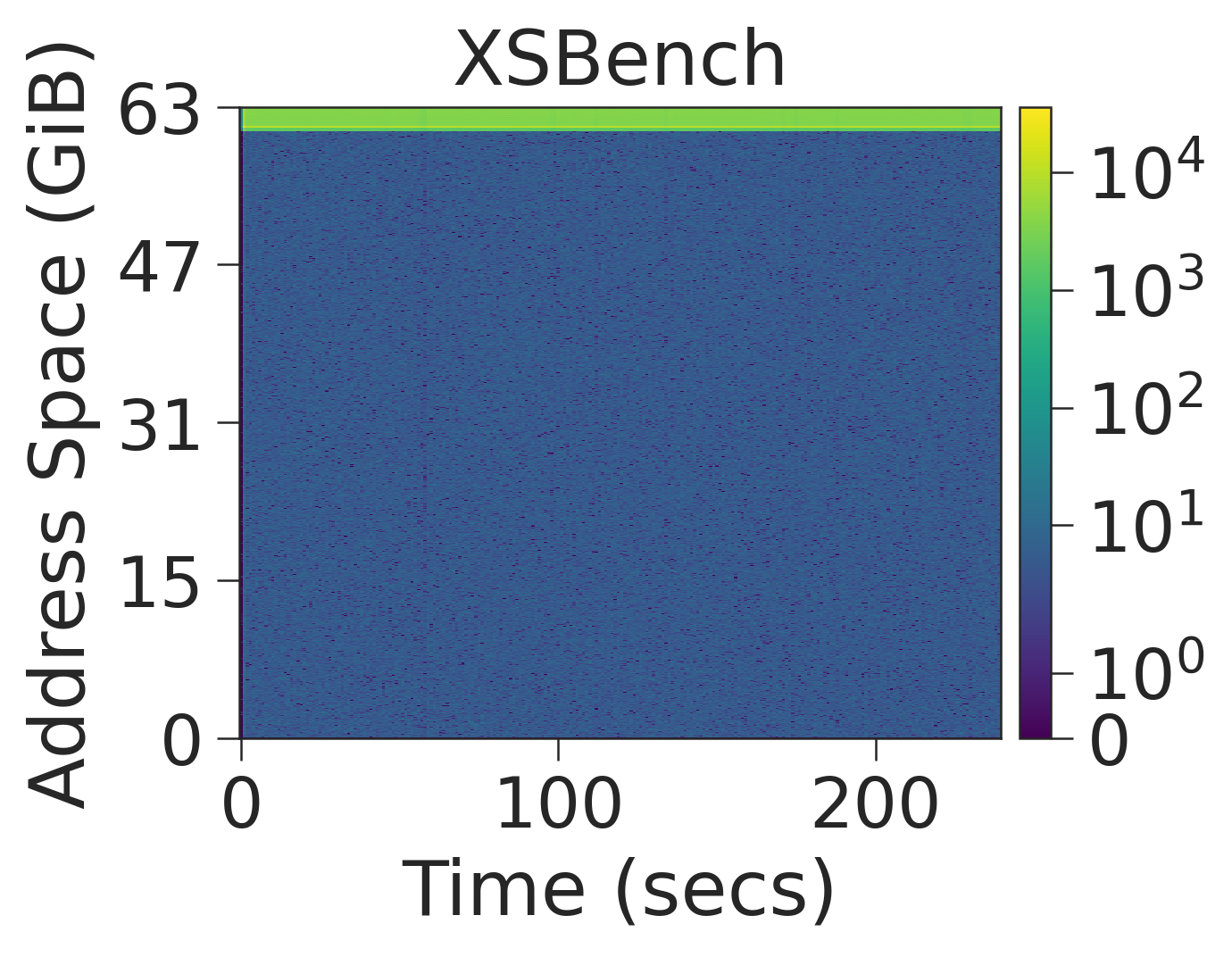}
    \end{minipage}
    \hfill
    \begin{minipage}{0.51\linewidth}
        \includegraphics[width=\linewidth]{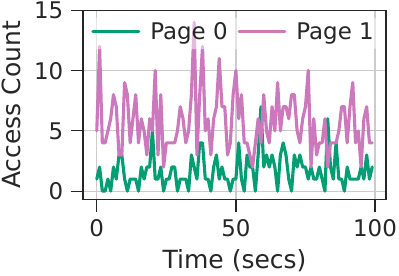}
    \end{minipage}
    \caption{\textit{XSBench}: Memory access pattern over time (left), and  page access count for two pages that are accessed uniformly and equally (right). Due to PEBS sampling inaccuracies, one page (i.e., Page 1) might appear hotter than another.}
    \label{fig:xsbench_pebs_pages}
    \vspace{-1ex}
\end{figure}

\textbf{3. Inability to handle access sampling noise and fluctuations:}
We observe that PEBS and DAMON introduce inaccuracies when sampling at low frequencies. For example, two pages accessed equally by the application may receive different sample counts over short time intervals. This is evident with XSBench that has the access pattern shown in Figure~\ref{fig:xsbench_pebs_pages}. Two adjacent pages in the region at the top (green) receive different samples.

Second, some pages experience a brief surge in popularity (ephemerally hot) but are not accessed afterward. Similarly, certain pages exhibit periodic access patterns. If the accesses count of these pages exceed the \texttt{hot\_threshold} at any point, existing systems classify these pages as hot and promote them. However, as their popularity fades, these pages quickly become cold and are subsequently demoted. This results in oscillations in the page migrations. A majority of these migrations are not useful, and degrade application performance by contending for memory bandwidth.

\vspace{-1ex}
\begin{mdframed}[linecolor=black,linewidth=1pt,innerleftmargin=8pt,innerrightmargin=8pt,innertopmargin=4pt,innerbottommargin=4pt, skipabove=8pt]
\textbf{Insight:} \textit{Tiering systems are not equipped to handle short-term noise or cyclical access oscillations, which results in wasteful migrations.}
\end{mdframed}

\subsection{Need for an adaptive and robust tiering engine}
\vspace{-.5ex}
The previous results reveal that existing engines are prone to making suboptimal decisions, mainly due to \textit{some} statically set parameter values.
Even systems like Memtis~\cite{memtis2023} that dynamically adjust some parameters underperform.

While tuning the parameters, e.g., using Bayesian Optimization, can greatly improve performance,
such techniques are typically very time consuming.
While prior work has attempted to identify a globally-best configuration for all workloads running on a fleet~\cite{googlezswap2019}, we find that no \textit{single} configuration can achieve best performance with existing systems. For example, not only does the best configuration for GapBS-BC perform poorly when used for Silo, the best configuration for GapBS-BC, can vary for different input graphs (e.g., \texttt{twitter} vs. \texttt{kronecker}).
This means that the tuning process has to be repeated for every application, input dataset and hardware platform, making it impractical for real world deployments.

To this end, we need a robust tiering engine, which can automatically adapt its behavior to the observed application and the underlying platform, avoiding the need for manual parameter tuning.
In the next section, we present our system, \sysname, which manages to meet the above goals.

\section{\sysname Design}
\label{sec:design}
\vspace{-.5ex}

Our goal is a tiering engine that delivers stable, high performance across workloads and platforms without per-workload tuning. 
To that end, we adopt four simple principles:
(i) \emph{prefer relative over absolute decisions:} rank pages by comparing against each other, rather than comparing with finicky thresholds to find the hot pages.
(ii) \emph{smooth inputs to eliminate noise:} convert raw page-access signals into stable estimates via lightweight smoothing/windowing so transient spikes do not trigger actions. 
(iii) \emph{adapt to system conditions:} utilize system state while making decisions. 
(iv) \emph{perform actions that have a positive expected value:} evaluate the benefit of a decision and act only when the expectation is positive.
These principles shape the design of \sysname.
 
\subsection{Design overview}
\vspace{-1ex}
\sysname overcomes the limitations of existing memory tiering systems using robust, adaptive policies and mechanisms. Below, we provide an overview of the key contributions:
\begin{enumerate}[topsep=6pt,leftmargin=12pt]
    \item \textbf{Ranking-based hotness:} \sysname uses a relative scoring algorithm using page access counts to identify the hot pages of a workload. \sysname computes a \textit{score} for every page based on its history of access. To strike a balance between accuracy and responsiveness, \sysname keeps access histories of different lengths: long and short. Long histories provide stability, overcoming short-term fluctuations whereas short histories help \sysname react quickly to changes in workload hot sets.
    \item \textbf{Adapting to workload hot set changes:} \sysname detects sudden changes in an application's hot set by monitoring changes in the application's slow tier accesses. On a sudden increase in accesses to the slow tier due to newly hot pages, \sysname utilizes the short access history to prioritize promotion of newly hot pages.
    \item \textbf{Filtering access bursts and fluctuations:} \sysname uses cost-benefit analysis to make migration decisions ensuring that all migrations are beneficial. Additionally it uses an adaptive observation window to filter pages with short bursts of accesses.
    \item \textbf{Adaptive batched migrations:} \sysname migrates pages concurrently, similar to Nimble~\cite{nimble2019}, to ensure that the hot pages are promoted as early as possible. In addition, \sysname ensures that migration bandwidth does not interfere with the application by dynamically adjusting the batch size. 
\end{enumerate}

\begin{figure}[t]
    \centering
    \includegraphics[width=\linewidth]{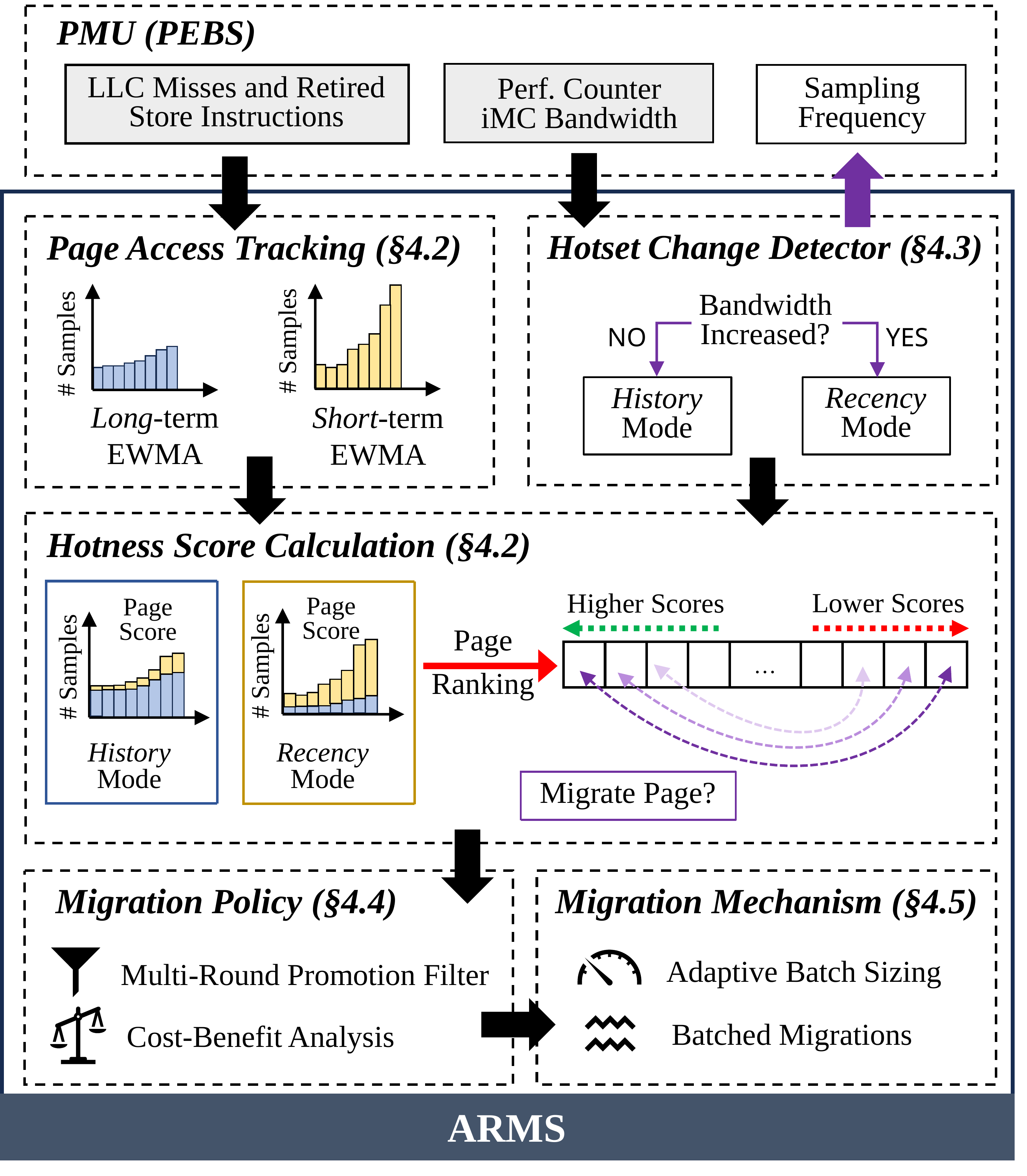}
    \caption{Architecture of the proposed system, \sysname.}
    \label{fig:arms-architecture}
\end{figure}

\noindent
Figure~\ref{fig:arms-architecture} provides an overview of \sysname design.
\sysname gathers memory reference and bandwidth information from hardware performance counters.
Periodically, \sysname updates the per-page access histories and detects if the application access distribution has changed.
With this new information, \sysname re-computes the scores for each page, and sorts them.
Finally, \sysname quickly identifies which pages need to be migrated, and moves them in parallel.

We now describe the details of \sysname design: tracking page access history, accurate page classification, adapting to workload changes, robust promotion policies and improved migration mechanism.

\subsection{Ranking-based hot page classification} 
Similar to prior tiering systems, \sysname uses page access history to predict future accesses and make page placement and migration decisions. \sysname uses hardware event sampling, such as Intel PEBS, to collect LLC load misses and retired store instructions. By default, \sysname uses a sampling rate of 1 in 10,000. We observe that sampling at this rate captures sufficient information about the workload without significant overhead. From our experiments, applications suffer about 0.3\% overhead from sampling at this frequency.

\sysname uses a relative scoring algorithm to identify the hottest pages of an application that should reside in the fast tier (DRAM) at any time, avoiding comparison to fixed thresholds. Second, to balance between accuracy and responsiveness, \sysname maintains two exponentially weighted moving averages (EWMAs) of the access counts for every page: a short-term $EWMA_{s}$ ($\alpha_{s}=0.7$) and a long-term $EWMA_{l}$ ($\alpha_{l}=0.1$). 
The short-term EWMA captures short-term trends which helps \sysname react quickly to changes in an application's hot set. The long-term EWMA adds robustness and captures historical trends, ignoring short-term fluctuations.
Because EWMAs discount the weight of older values, \sysname does not perform periodic cooling.

\mypar{Page scoring:}
\sysname combines the two EWMAs to calculate a \textit{hotness score} for every page periodically. It then ranks application pages by this hotness score. The top-k pages, where \textit{k} is the capacity of the fast tier, are considered hot and chosen to be promoted if they are not already in the fast tier. This ranking algorithm has 2 benefits: it ensures that the desired number of hot pages are always identified, and enables pages with the most number of accesses to be prioritized for promotion.

The hotness score is a weighted sum of the two EWMAs. \sysname adapts the weights to the application state/phase. When the application is in a steady state with no change in the hot set, \sysname prioritizes the long-term average over the short-term average. \sysname therefore avoids promoting pages with transient accesses and ensures that historically hot pages are ranked higher. However, when there is a change in the workload access distribution, \sysname increases $w_{s}$, weight of the short-term average, i.e., \sysname prioritizes recent access history over long-term history. This enables \sysname to identify and promote the new hot pages quickly.

\begin{mdframed}[linecolor=black,linewidth=1pt,innerleftmargin=8pt,innerrightmargin=8pt,innertopmargin=4pt,innerbottommargin=4pt, skipabove=8pt]
\textbf{Summary:} \textit{\sysname maintains moving averages of different lengths to capture both historical and recent access patterns. It combines the averages using dynamic weights to calculate page scores which is then used to rank pages.}
\end{mdframed}

\subsection{Adapting score weights based on workload state}
For good performance, tiering systems need to react quickly to changes in the application hot set, especially when the new set of hot pages are mainly in the slow tier. As we discussed in Section~\ref{sec:tuning_analysis}, existing systems can be slow to react to such scenarios, and hence suffer from poor performance.
\sysname detects sudden changes in an application's hot set using a change point detection algorithm based on the \textit{Page-Hinkley test} (PHT), which is a sequential analysis technique used to monitor changes in data~\cite{pht1954}.

Our key insight is that when working sets change, performance can suffer if more memory references go to the slow tier. In other words, if there are a new set of hot pages in the slow tier, then there is an increase in the number of accesses to the slow tier.

\sysname uses hardware performance counters to monitor an application's accesses to slow tiers, and employs the PHT algorithm to detect any increase in the slow tier bandwidth utilization.
Upon detecting such a change, \sysname enters \textit{recency mode}. In this mode, as mentioned above, \sysname increases the weights (prioritizes) of short-term EWMAs and de-prioritizes long-term EMWAs for the hotness score calculation. This allows \sysname to overlook this historical hotness of pages and pay attention to the recent trends, making it more responsive than prior systems.

In addition, in recency mode, \sysname doubles the PEBS sampling rate to gather more accurate information about the new workload behavior. The system stays in this mode for a fixed period or until the bandwidth utilization has stabilized. Once new hot pages are promoted and the application reaches a stable execution phase, \sysname returns to the default \textit{history mode}, where it prioritizes the long-term EWMAs and reduces the sampling frequency.

\vspace{-1ex}
\begin{mdframed}[linecolor=black,linewidth=1pt,innerleftmargin=8pt,innerrightmargin=8pt,innertopmargin=4pt,innerbottommargin=4pt, skipabove=8pt]
\textbf{Summary:} \textit{\sysname detects changes in hot set by monitoring slow tier access rate of the application. If a change is detected, \sysname adjusts the EWMAs weights to prioritize recent access history, promoting new hot pages early.}
\end{mdframed}

\begin{algorithm}[t]
\footnotesize
\begin{algorithmic}
\State $k$: Fast tier capacity
\State $Pages_{all}$: list of application pages
\State $Pages_{topk}$: top k hot pages in current interval
\State $w_l, w_s$: score weights adapted based on recency or history mode
\end{algorithmic}

\begin{algorithmic}[1]
\Statex
\State // \textcolor{blue}{Hotness score calculation}
\For{$P$ in $Pages_{all}$}
    \State $P_{EWMA_{s}} = \alpha_{s} \times P_{EWMA_{s}} + (1-\alpha_{s}) P_{accesses}$
    \State $P_{EWMA_{l}} = \alpha_{l} \times P_{EWMA_{l}} + (1-\alpha_{l}) P_{accesses}$
    \State $P_{prevscore} = P_{score}$
    \State $P_{score} = w_{s} \times P_{EWMA_{s}} + w_{l} P_{EWMA_{l}}$
\EndFor
\Statex
\State // \textcolor{blue}{Identifying top k hot pages}
\State $Pages_{sorted}$ = \textbf{Sort} $Pages_{all}$ in descending order of $P_{score}$
\State $Pages_{topk}$ = $Pages_{sorted}[0:k-1]$

\end{algorithmic}
\caption{Ranking-based hotness classification}
\label{alg:page_classification}
\end{algorithm}

\subsection{Avoiding migrations due to access bursts and oscillations}
\label{sec:wasteful_migrations}

PEBS sampling can have significant inaccuracies (see Section~\ref{sec:tuning_analysis}), causing some pages to have a sudden bursts of samples. Additionally, some workloads exhibit pathological behaviors wherein some pages stay hot for an extremely short duration~\cite{memtis2023,flexmem2024}.
Existing systems do not handle these scenarios well which leads to migration churn which degrade application performance.
\sysname incorporates two strategies to prevent wasteful migrations:

\mypar{Cost-benefit analysis of migrations.}
Inspired by prior work on cost-benefit models for huge page management~\cite{cbmm2022}, \sysname ensures that all migrations lead to performance gains. \sysname promotes a page only if its estimated benefit exceeds the associated migration cost.

\textbf{1) Estimating cost:} Page migrations stall the application if there is a racing write to the migrating page. In the worst case, an application could stall for the entire duration of migration. In \sysname, we take a conservative approach and assume the worst case cost for every migration. \sysname uses the measured latencies of past promotions and demotions to estimate the cost of a future migration.

\textbf{2) Estimating benefit:}
To quantify the benefit of a promotion, \sysname estimates the potential improvement in application performance due to the reduced memory access latencies if the page resides in the fast tier. For this, \sysname estimates the number of future accesses to the page. It uses (1) the page's hotness score and (2) the predicted hot age (time for which the page will stay hot). \sysname predicts the hot age based on the historical data for the application. It tracks the average time pages stay hot, and uses this to estimate the future hot age for any page.
Second, \sysname estimates the latency difference between the two tiers based on the current bandwidth. 
We run an offline analysis to capture the bandwidth-latency curve for each memory tier. We use a piecewise RELU-like model wherein the memory latency is constant ($L_{0}$) till the bandwidth knee point ($B_{k}$), then increases linearly ($m$).
Then during execution, the latency at a specific bandwidth ($L(B)$) is estimated using the measured bandwidth (see Algorithm~\ref{alg:promotion_criteria} for details).

By ensuring that the performance gains from reduced memory access latencies outweigh the migration overhead, this cost-benefit approach helps avoid any unnecessary migrations and improves overall system efficiency.

\mypar{Multi-round promotion filtering.}
To prevent promotion of ephemeral hot pages (pages which are hot for a very short duration) \sysname uses a promotion filtering strategy that ensures that only pages that continue to stay hot are promoted.
When a page gets a high hotness score and enters the set of top-k hot pages, \sysname does not promote the page instantly. Instead, it marks the page as a candidate for promotion and continues to monitor the page.
\sysname adjusts the monitoring period based on the page score. A page with a sufficiently high score is monitored for a single period and promoted early, while a page with a lower score is monitored for longer, upto $N_{max}$ periods, to ensure the page continues to stay hot. 
If the page continues to stay in the top-k hot pages and its score continues to increase or stay the same, then \sysname promotes it. If the page drops out of the top k or its score goes down, it is not promoted. 

\vspace{-1ex}
\begin{mdframed}[linecolor=black,linewidth=1pt,innerleftmargin=8pt,innerrightmargin=8pt,innertopmargin=4pt,innerbottommargin=4pt, skipabove=8pt]
\textbf{Summary:} \textit{\sysname filters migrations using a cost-benefit estimation and a hot-persistence test. This way, it avoids any wasteful migrations due to access sampling inaccuracies or short burts of accesses.}
\end{mdframed}

\begin{algorithm}[t]
\footnotesize
\begin{algorithmic}
\State \textcolor{blue}{// Variables pre-calculated offline}
\State $L_{fast_0}, L_{slow_0}$: fast and slow tier unloaded latency
\State $BW_{fast_k}, BW_{slow_k}$: fast and slow tier bandwidth knee point
\State $m_{fast}, m_{slow}$: fast and slow tier bandwidth slope
\State \textcolor{blue}{// Variables computed online}
\State $P_{topk}$: list of top k hot pages
\State $P_{candidates}$: list of promotion candidates
\State $P_{cold}$: list of cold pages
\State \textcolor{blue}{// Parameters}
\State $N_{max}$: max monitoring windows

\end{algorithmic}
\begin{algorithmic}[1]

\Statex
\State \textcolor{blue}{// Multi-round promotion filtering}
\For{$p$ in $P_{topk}$}
    \State $r = (p_{score} - min\_score) / (max\_score - min\_score)$
    \State $N = 1 + (1-r) * (N_{max} - 1)$  \textcolor{blue}{// Monitoring window}
    \If{$p_{score} >= p_{prevscore}$ \textbf{and} $p_{hotage} >= N$}
        \State $P_{candidates} \gets P_{candidates} \cup \{p\}$
    \EndIf
\EndFor

\Statex
\State \textcolor{blue}{// Cost-benefit analysis}
\State $L_{fast} = L_{fast_0} + m_{fast}(BW_{fast} - BW_{fast_k})$
\State $L_{slow} = L_{slow_0} + m_{slow}(BW_{slow} - BW_{slow_k})$
\State $\Delta L = L_{slow} - L_{fast}$
\State $Cost = L_{promotion} + L_{demotion}$
\For {$p$ in $P_{candidates}$}
    \State $q \gets \text{coldest page from } P_{cold}$
    \State $Benefit = (p_{score}-q_{score}) \times p_{hotage} \times \Delta L$
    \If{$Benefit > Cost$} \textbf{Demote} $q$ and \textbf{Promote} $p$
    \EndIf
\EndFor

\end{algorithmic}
\caption{Promotion criteria}
\label{alg:promotion_criteria}
\end{algorithm}

\subsection{Improved migration mechanism}

Hot pages that meet the promotion criteria described above are chosen to be promoted. Instead of promoting hot pages in the order they are identified, \sysname employs a priority-based scheduling mechanism, where priority is determined by the hotness score. This ensures that the hottest pages are promoted first and do not suffer from head-of-line blocking observed in prior systems~\cite{hemem2021}.

Second, \sysname leverages the free bandwidth available in the system to perform batched migrations, similar to Nimble~\cite{nimble2019}, thereby reducing the promotion delay for hot pages.
To prevent the increased migration bandwidth from degrading application performance, \sysname dynamically adjusts the batch size $BS$ based on the application's bandwidth usage:

\begin{equation*}
    BS = max(1,\frac{(BW_{k} - BW_{app})}{BW_{k}} * BS_{max})
\end{equation*}
\vspace{1ex}

To determine $BS_{max}$, we conduct offline analysis to identify the number of threads required to saturate the bandwidth. \sysname measures the real-time application bandwidth using hardware performance counters as described earlier. By dynamically sizing the migration concurrency, \sysname ensures that pages are migrated as early as possible without adversely affecting application performance. For instance, during brief periods of application inactivity, \sysname maximizes the number of migrations to fully exploit the available system bandwidth while maintaining performance stability.

\section{Implementation}
\label{sec:implementation}

We implement \sysname as a user-level library, similar to HeMem.
Applications link to the library using \texttt{LD\_PRELOAD}. The library intercepts memory allocation calls such as \texttt{mmap} and \texttt{free} and handles memory allocation and freeing of anonymous regions. Each memory tier is exposed as a DAX (direct-access) file, which is mapped by the library during startup. We adopt HeMem's approach to forward page faults to user-space using userfaultfd during which the library allocates pages from the mapped DAX files. 

\sysname employs hugepages and tracks page access counts and all metadata at 2MB  granularity. Huge pages  not only reduce address translation overheads but also reduces  computational and memory costs. \sysname maintains an access counter, two averages, two scores, and a hot age per page. This adds about 20 bytes per page. At hugepage (2MB) granularity, this results in negligible memory overhead. For instance, tracking 200 GB of pages incurs only a 2 MB overhead (0.001\% overhead).
 
\mypar{PEBS sampling thread.} \sysname uses Intel PEBS to sample LLC load misses and retired store instructions.  \sysname switches between sampling periods of 10K (default) and 5K (recency mode).
The library spawns a dedicated thread to process  PEBS samples and update  access counts. Our current implementation polls continuously on a dedicated core. This could be  optimized to poll periodically, similar to Memtis~\cite{memtis2023}, which would reduce the CPU usage to a single digits (3\% in Memtis).

\mypar{Policy thread.} The library  spawns a policy thread that periodically updates page hotness scores and make migration decisions. The period is set similarly to the PEBS sampling rate. During steady phases, this thread runs every 500ms but when \sysname detects a hot set change, it runs every 100ms to ensure that hot pages are identified and promoted quickly. On average, this thread consumes 8.6\% of a single CPU core. 

\section{Discussion}
\label{sec:discussion}

\begin{table}[t]
\centering
\resizebox{\linewidth}{!}{%
\begin{tabular}{lll}
\toprule
\textbf{Mechanism} & \textbf{Parameter} & \textbf{Description} \\
\midrule
\begin{tabular}[t]{@{}l@{}}Hotness\\ scoring\end{tabular} &
\begin{tabular}[t]{@{}l@{}}
\textbf{\texttt{sampling\_period}} \\
\texttt{ewma\_short\_window} \\
\texttt{ewma\_long\_window} \\
\textbf{\texttt{w\_short}, \texttt{w\_long}} \\
\end{tabular} &
\begin{tabular}[t]{@{}l@{}}
Access sampling interval \\
Short-term moving averaging window \\
Long-term moving averaging window \\
Weights for combining EWMAs into score
\end{tabular} \\
\midrule
\begin{tabular}[t]{@{}l@{}}Hot-set change\\ detection\end{tabular} &
\begin{tabular}[t]{@{}l@{}}
\texttt{bw\_avg\_window} \\
\textbf{\texttt{hcd\_trigger}} \\
\textbf{\texttt{hcd\_drift}} \\
\texttt{recn\_min\_duration} \\
\end{tabular} &
\begin{tabular}[t]{@{}l@{}}
Slow tier bandwidth averaging window \\
Change detection threshold \\
Tolerance to prevent overreactions \\
Minimum duration to stay in recency mode \\
\end{tabular} \\
\midrule
\begin{tabular}[t]{@{}l@{}}Cost-benefit\\ analysis\end{tabular} &
\begin{tabular}[t]{@{}l@{}}
\texttt{cost\_avg\_window} \\
\end{tabular} &
\begin{tabular}[t]{@{}l@{}}
Migration cost averaging window
\end{tabular} \\
\midrule
Promotion filter &
\begin{tabular}[t]{@{}l@{}}
\textbf{\texttt{max\_monitor\_time}}
\end{tabular} &
\begin{tabular}[t]{@{}l@{}}
Maximum time to monitor for hotness
\end{tabular} \\
\bottomrule
\end{tabular}%
}
\vspace{.5ex}
\caption{\sysname parameters and thresholds. Parameters in bold are dynamically adjusted based on observed workload characteristics and system state, others are static.}
\label{tab:system_params}
\end{table}

Below, we discuss some design choices and future work.

\mypar{\sysname parameters.} \sysname eliminates sensitive thresholds used in prior systems but does use some internal parameters shown in Table~\ref{tab:system_params}.
Most of these parameters are dynamically adjusted based on the workload behavior and the measured system state.
From our experiments, we observe that applications are insensitive to these parameters (see Section~\ref{sec:eval:sensitivity}), and a single set of values work well in all scenarios, making \sysname robust.

\mypar{Optimizations in prior work.} Recent works have proposed many optimizations to tiering systems. Nomad~\cite{nomad2024} introduces a transactional page migration mechanism which decouples data movement from application accesses. This can be combined with \sysname's migration mechanism.

Colloid~\cite{colloid2024} argues that tiering systems should consider access latencies while making tiering decisions. Colloid was implemented on top of HeMem, Memtis and TPP, and is also applicable to \sysname. The key contribution of \sysname is accurate and timely identification of hot and cold pages. With Colloid, \sysname would incorporate an additional check based on access latencies for page migrations.

Memtis~\cite{memtis2023}, TPP~\cite{tpp2023} and Flexmem~\cite{flexmem2024} maintain a pool of free memory in the fast tier for faster promotions of hot pages. This is complimentary to the contributions of \sysname.

\mypar{Reducing overheads.} Our evaluation shows that the benefits of \sysname outweigh its computation overheads. They can be lowered by reducing the periodicity of execution of page classification. Currently, the policy thread runs every 500ms in steady phases, but could run pause if there are no migrations and be woken by PHT when it detects a change in application behavior. 

\section{Evaluation}
\label{sec:eval}

\begin{table}[t]
\small
\centering
\begin{tabular}{@{}lccc@{}}
\toprule
\textbf{Specification} & \textbf{NUMA} & \textbf{pmem} \\ \midrule
Number of cores                   & 20      & 24     \\
Processor generation               & Skylake & Icelake \\
Processor frequency (GHz)          & 2.2       & 3    \\
L3 cache size (MB)                 & 13.75      & 18 \\
\hline
Far memory type                    & NUMA  & Optane   \\
Max near memory size (GB)          & 96      & 96     \\
Max far memory size (GB)           & 96     & 128     \\
Max near mem BW (GB/s)             & 56     & 138     \\
Max far mem BW R/W (GB/s)          & 36/36  & 7.45/2.25 \\
Near memory latency (ns)   & 95      & 80     \\
Far memory latency (ns)    & 180 &  200 \\
\bottomrule
\end{tabular}
\vspace{1ex}
\caption{Specifications of machines used in our evaluation.}
\label{tab:machine_spec}
\end{table}

Through our experiments, we answer the following questions about \sysname:
\begin{enumerate}[leftmargin=0.5cm,topsep=0pt]
    \item What are the performance benefits of \sysname compared to state-of-the-art tiering systems and tuned systems?
    \item How do \sysname's new policies and mechanisms perform?
    \item Is \sysname robust and adaptive? Does it work well on different hardware, different memory configurations and different application inputs?
    \item Do \sysname parameters require tuning?
\end{enumerate}

\myparnospace{Hardware setup.}
Table~\ref{tab:machine_spec} shows the hardware specifications of the two machines used for our experiments. 
\texttt{NUMA} machine (provided by Cloudlab~\cite{cloudlab}) uses remote NUMA node to emulate CXL memory, whereas \texttt{pmem} uses Intel Optane DC Persistent Memory DIMMs as the slow tier.
In this section, we primarily share results from experiments on \texttt{NUMA}, unless mentioned otherwise.

\myparnospace{Workloads.}
We select a set of nine representative and diverse workloads: a classifier training kernel from Liblinear~\cite{fan2008liblinear}, three graph processing algorithms from GapBS~\cite{beamer2015gap}, an HPC workload (XSBench \cite{tramm2014xsbench}), two in-memory database workloads (Silo \cite{tu2013speedy} with TPC-C and YCSB), an in-memory index lookup (Btree \cite{reto2020btree}), and a popular memory-intensive micro-benchmark, GUPS \cite{plimpton2006simple}. 
Table~\ref{tab:benchmark_specs} contains additional information about these workloads.
These workloads have been widely used to evaluate tiered memory systems in many prior works \cite{hemem2021, memtis2023, johnnyCache2023,autotiering2017}.
We run each application with 16 threads, which is enough to just saturate the memory bandwidth of each system.

\begin{figure*}
    \centering
    \includegraphics[width=\linewidth]{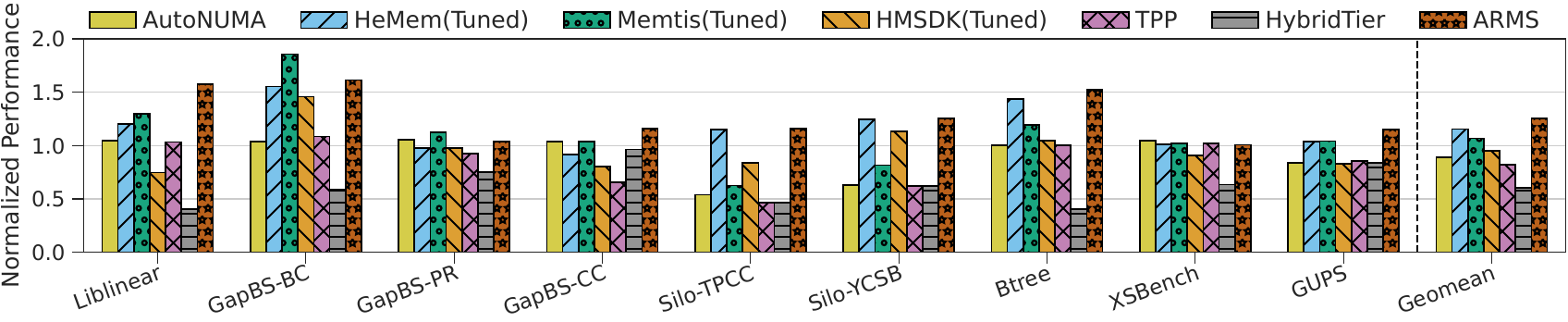}
    \caption{Comparison of \sysname with existing systems on \texttt{NUMA} machine. HeMem, Memtis, and HMSDK use the best-performing configuration for each workload; other systems use their default. Performance normalized to static allocation. Higher is better.}
    \label{fig:perf_comparison_numa}
    \vspace{-2ex}
\end{figure*}

\myparnospace{Tiering Configuration.} Similar to Memtis~\cite{memtis2023}, we configure the ratio of fast-to-slow tier memory size by setting the fast tier size to a fraction of the workload resident set size (RSS). Unless otherwise noted, experiments maintain a \texttt{1:8} fast:slow memory size ratio. 

\begin{table}[t]
\scriptsize
\centering
\begin{tabular}{@{}llll@{}}
\toprule
\textbf{Workload} & \textbf{Inputs} & \textbf{RSS} & \textbf{Description} \\ \midrule
Liblinear \cite{fan2008liblinear} & kddb & 34.13 &  Logistic regression-based classifier \\ \midrule
\multirow{2}{*}{GapBS-BC \cite{beamer2015gap}} & kronecker  & 78.13 & Compute the measure of centrality  \\
 & twitter & 13.08 &  in a graph based on shortest paths. \\ \midrule
\multirow{2}{*}{GapBS-PR \cite{beamer2015gap}} & kronecker  & 71.29 & \multirow{2}{*}{Compute the PageRank score of a graph} \\
 & twitter & 12.32 & \\ \midrule
\multirow{2}{*}{GapBS-CC \cite{beamer2015gap}} & kronecker  & 69.29 & Compute connected components of \\
 & twitter & 12.09 &  a graph using (Shiloach-Vishkin) \\ \midrule
\multirow{2}{*}{Silo \cite{tu2013speedy}} & TPC-C & 75.68 & \multirow{2}{*}{In-memory transactional database} \\
 & YCSB-C & 71.40 & \\ \midrule
Btree \cite{reto2020btree} & - & 12.13 &  In-memory index lookup benchmark \\ \midrule
XSBench \cite{tramm2014xsbench} & - & 64.97 & Compute kernel of the MCNP algorithm \\ \midrule
GUPS \cite{plimpton2006simple} & 8 GiB hot & 64.03 & Random accesses with dynamic hotset \\
\bottomrule
\end{tabular}
\vspace{1ex}
\caption{Workload Characteristics. RSS is in GiB.}
\label{tab:benchmark_specs}
\end{table}

\myparnospace{Systems compared.} We compare \sysname with:

\begin{enumerate}[topsep=2pt,leftmargin=.5cm]
    \item Static partitioning: No migrations after initial allocations.
    \item HeMem~\cite{hemem2021}: We compare against both default HeMem and tuned-HeMem.
    \item Memtis~\cite{memtis2023}: We compare against both default and tuned-Memtis with Linux v5.15. We disable huge page splitting.
    \item HMSDK~\cite{hmsdk2024}: We compare against both default and tuned-HMSDK with Linux v6.12.
    \item AutoNUMA~\cite{autonuma}: We use Linux AutoNUMA v6.2. We ensure demotion is enabled and toptier nodes are set correctly.
    \item TPP~\cite{tpp2023}: We use TPP with Linux v6.0. We apply the promotion policies not up-streamed to Linux yet.
    \item HybridTier~\cite{hybridtier2025}: We use HybridTier with Linux v6.2.
\end{enumerate}
We enable hugepages for all systems. We obtain the tuned configurations using the method discussed in Section~\ref{sec:tuning_study}.

\subsection{Performance comparison}

Figure~\ref{fig:perf_comparison_scailp} shows the performance of \sysname and other tiering systems normalized to static partitioning. Without tuning, \sysname matches or outperforms tuned HeMem, Memtis and HMSDK. On average (geomean), \sysname performs $1.22\times$ ($1.05\times$) better than default (tuned) HeMem, $1.23\times$ ($1.1\times$) better than default (tuned) Memtis, $1.33\times$ better than TPP, $1.28\times$ better than AutoNUMA, and $1.85\times$ better than HybridTier.

\mypar{Comparison to HeMem.} Untuned HeMem performs poorly because of the reasons discussed in Section~\ref{sec:tuning_analysis}: (1) inaccurate hot pages identification, (2) hot page promotion delays, and  (3) wasteful migrations.
Figure~\ref{fig:wasteful_migrations} compares the number of migrations performed by different tiering systems. When running XSBench, HeMem performs a large number of wasteful migrations which consumes valuable memory bandwidth and degrades application performance because it is unable to handle the PEBS sampling inaccuracies and fluctuations (see Figure~\ref{fig:xsbench_pebs_pages}).
The migration cost-benefit analysis and multi-round promotion filtering in \sysname provides immunity against such inaccuracies. While tuned-HeMem overcomes the above problems, \sysname is able to achieve the same performance without requiring expensive tuning.

\mypar{Comparison to Memtis.} In spite of the superior hot page classification of Memtis, we find that untuned Memtis performs poorly for many workloads, especially those with hot set changes.
One of the main reasons for this is \textit{inaccurate cooling threshold}. Memtis performs cooling for every two million PEBS samples collected. 
Consider TPC-C, which is write-heavy and follows a "latest distribution".
Memtis samples writes at an extremely low frequency, so cooling is performed once every ~100s.
Unfortunately, due to infrequent cooling, Memtis overestimates the hot pages and is not able to promote recently hot pages.
On the other hand, \sysname does not use a static threshold for cooling, it adapts to changes in the hot set and promotes new hot pages quickly.

\mypar{Comparison to AutoNUMA and TPP.} AutoNUMA and TPP achieve lower performance than most systems. Both AutoNUMA and TPP rely on page faults to detect hot pages. AutoNUMA promotes on the first hint fault whereas TPP promotes if it is faulted twice. Since these systems rely on access recency more than frequency, we observe that these systems are unable to differentiate between hot pages and warm pages. When the set of hot + warm pages exceeds the fast tier, these systems suffer from high migration churn as shown in Figure~\ref{fig:wasteful_migrations}.
Second, we also observe a high number of promotion failures~\cite{flexmem2024,pet2025} because the promotion and demotion paths are decoupled. This delays promotion of hot pages.
While these policies could work well in systems with larger fast tier capacities (2:1 ratio), they are insufficient on machines with more skewed ratios. In comparison, \sysname tracks both access frequency and recency information, leading to better placement decisions.

\mypar{Comparison to HybridTier.} Similar to \sysname, HybridTier also uses two averages called \textit{frequency} and \textit{momentum} to determine page hotness. Surprisingly, we find that HybridTier has the least performance.
Unlike \sysname, HybridTier uses thresholds to classify hot and cold pages. For example, the momentum threshold is set to 300 (for hugepages) which works well for some workloads like GUPS but does fails for workloads like Silo-TPCC. We observe that this results in inaccurate hot page identification. With Silo-YCSB which has a fixed set of hot pages, the number of hot pages classified by HybridTier keeps fluctuating. Sometimes, it identifies only 500 hot pages, while at other times it classifies as many at 12000, when the DRAM capacity is around 4000 pages.
The bigger problem, however, is HybridTier's cold page demotion. It scans process page tables, similar to DAMON, to identify cold pages. After candidate cold pages are identified, Hybridtier waits an additional minute before demoting the cold pages. As a result, we observe long promotion delays.

\subsection{Understanding \sysname performance}

\begin{figure}
    \centering
    \setlength{\abovecaptionskip}{4pt}
    \includegraphics[width=\linewidth]{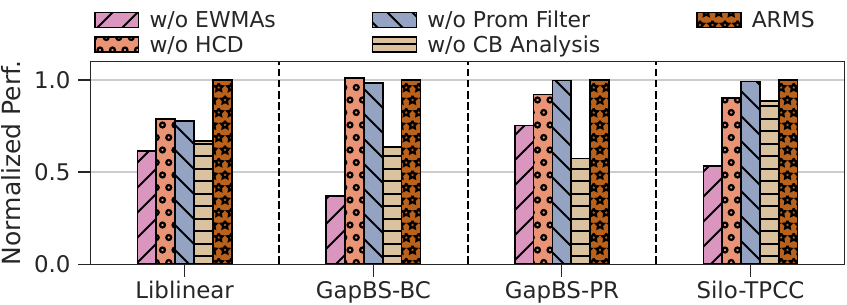}
    \caption{Performance when \textit{not} using (1) EMWA averaging and scoring, (2) hot-change detector (HCD), (3) promotion filter, and (4) cost-benefit analysis. Performance normalized to \sysname. Higher is better.}
    \label{fig:breakdown-fig}
\end{figure}

\sysname outperforms existing tiering systems through the use of adaptive, robust policies for identifying hot pages and migrating them in a timely manner. Here, we discuss the effectiveness of the different contributions of \sysname.
Figure~\ref{fig:breakdown-fig} shows the impact of different components of \sysname on application performance. We now focus on each component.

\mypar{Ranking-based hotness classification.} A key driver of \sysname’s performance is more accurate hot-page detection. Both ranking and the use of averages are key contributors to performance. As Figure~\ref{fig:breakdown-fig} shows, relying on raw instantaneous counts instead of averages hurts performance. Averaging smooths noisy PEBS samples, yielding more reliable hotness estimates. while ranking selects the capacity-matched top-$k$ pages rather than chasing threshold crossings. Together, these choices place the \textit{truly} hot pages in fast tier.

\begin{figure}
\centering
\setlength{\abovecaptionskip}{4pt}
\includegraphics[width=\linewidth]{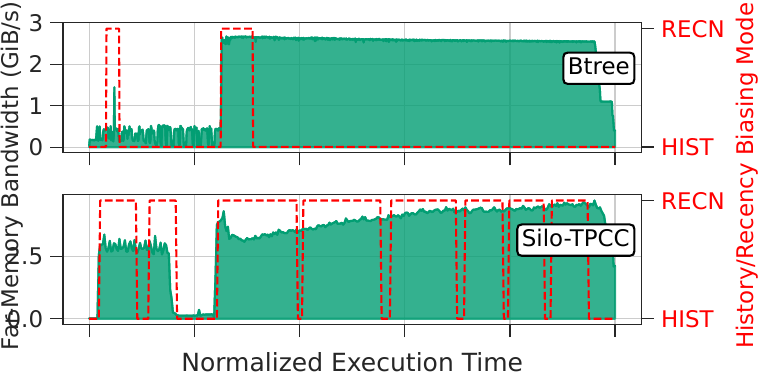}
\caption{Slow tier bandwidth usage (left y-axis) and hot set change detection mechanism (right y-axis) timeline, for \textit{Btree} and \textit{Silo-TPCC}. By default, \sysname uses history (\texttt{HIST}) mode, prioritizing historical data. When an \textit{increase} in far-memory bandwidth is detected, indicating a change in hot-set, \sysname switches to recency (\texttt{RECN}) mode, prioritizing recent samples.}
\label{fig:hotsetchangedetection}
\vspace{2ex}
\end{figure}

\mypar{Adapting to changes in application access pattern.} 
Figure~\ref{fig:hotsetchangedetection} shows the slow tier bandwidth change detection in action for two workloads. Whenever it detects an increase in the accesses to slow tier, \sysname transitions into recency mode to prioritize recent access history.
The detector's behavior with Silo-TPCC is interesting. As mentioned previously, TPC-C follows a recent distribution wherein newly inserted tuples remain hot and get cooler over time. As new elements are added and the database grows larger, the traffic to the slow tier also increases. The hot set change detector triggers continuously as seen in Figure~\ref{fig:hotsetchangedetection} and keeps \sysname in recency mode for most of the time. \sysname is therefore able to prioritize recent history and keep the new pages in fast tier.

Our experiments demonstrate that the algorithm accurately detects most hot set changes. While the algorithm occasionally triggers false positives (<2 on average per workload), we observe that these misclassifications have negligible performance impact. Figure~\ref{fig:breakdown-fig} shows that the hot set change detector is important for workloads with changing access distribution such as Liblinear and Silo-TPCC.

\begin{figure}
    \centering
    \setlength{\abovecaptionskip}{4pt}
    \includegraphics[width=\linewidth]{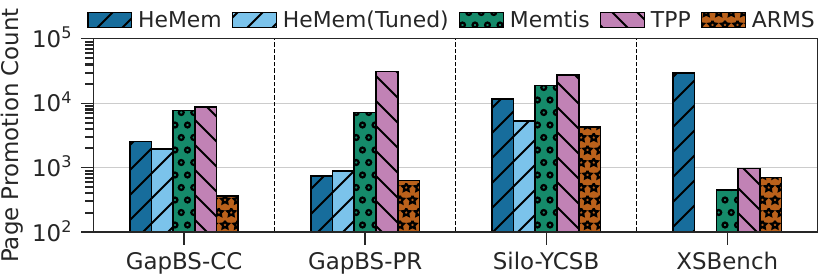}
    \caption{Page promotion count for HeMem, Memtis, TPP and \sysname. Y-axis is in log scale. Lower is better.}
    \label{fig:wasteful_migrations}
    \vspace{-1ex}
\end{figure}

\mypar{Minimizing wasteful migrations.} The multi-round promotion filtering and cost-benefit analysis ensure that \sysname only performs migrations that are beneficial. From Figure~\ref{fig:wasteful_migrations}, we observe that \sysname makes the least number of migrations for many workloads, and achieves the best performance. Other systems are twitchy, react eagerly to spikes and fluctuations, leading to a large number of wasteful migrations.

We observe that the cost-benefit analysis is especially useful for warm pages. The cost-benefit analysis prevents of migrations of warm pages as the potential benefit is minimal. Prior systems~\cite{memtis2023, flexmem2024} try to achieve this by classifying warm pages and preventing their migration. They still suffer from wasteful migrations because of incorrect thresholds. 

\mypar{Dynamic batched migrations.} To understand the benefit of adaptive batched migrations, we compare \sysname behavior with and without batching. We find that applications like Liblinear that have periodic phases benefit the most as many pages can be migrated during non-memory-intensive phases. For Liblinear, adaptive batched migrations provide about 7\% performance improvement, and $\sim$2\% for CC, PR and GUPS.

\subsection{\sysname parameter stability}
\label{sec:eval:sensitivity}
\begin{figure}[t]
    \centering
    \setlength{\abovecaptionskip}{4pt}
    \begin{minipage}{0.48\linewidth}
    \includegraphics[width=\linewidth]{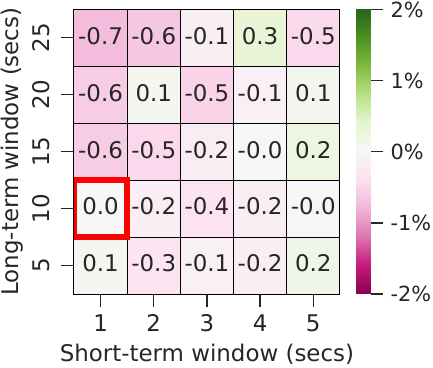}
    \end{minipage}
    \hfill
    \begin{minipage}{0.48\linewidth}
        \includegraphics[width=\linewidth]{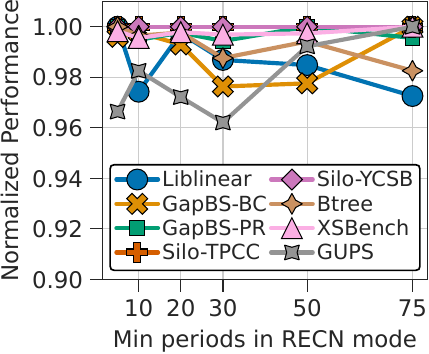}
    \end{minipage}
    \caption{Performance variability of \sysname when tweaking short/long-term EMWAs window length (left), and when tweaking the number of periods to stay in \texttt{RECN} mode after switching from \texttt{HIST} mode (right). Performance normalized to \sysname's default parameter values.}
    \label{fig:arms-sensitivity}
\end{figure}

One of the main goals of \sysname is to use robust policies that avoid application performance sensitivity to tiering parameters.
Most of \sysname parameters, shown in Table~\ref{tab:system_params} are adjusted dynamically based on the workload behavior and system state.
For the remaining \textit{static} ones, we conduct a sensitivity study.
Figure~\ref{fig:arms-sensitivity} (left) shows the performance variation when the short-term and long-term moving average windows are changed. The numbers in the boxes show the geomean of all applications performance (standard deviation is $<0.5\%$). We observe that the window lengths have minimal impact on performance. This suggests that one can set the window length to any reasonable value and forget.

We also study the effect of \texttt{recn\_min\_duration} parameter which decides how long \sysname stays in recency mode if the hot set change detector identifies a change in the access distribution.
The performance of most workloads is unaffected by this parameter. For the other workloads, the variation is less than 4\%. 
Staying in recency mode for long only affects performance if there are lot of fluctuations in the input access samples. \sysname, however, increases the sampling frequency in recency mode to overcome the fluctuations. Therefore, the penalty for mis-predictions or staying longer in recency mode is very minimal.
One way to avoid this parameter altogether is to detect if all the new hot pages have been promoted and return to history mode. However, we have not implemented this yet.

We perform similar studies for the remaining knobs and observe very little to none performance variation.

\subsection{Adaptability to different scenarios}
We demonstrate that \sysname works well on a different machine, with different thread counts, and with different fast:slow memory ratios. We do not modify any \sysname parameter but run offline analysis to capture the bandwidth-latency curve, and $BS_{max}$ of the new platform. For these studies, we cannot compare against HeMem-tuned due to the high cost of tuning for every application and configuration. We discuss each of these results below.

\begin{figure}
    \centering
    \includegraphics[width=0.95\linewidth]{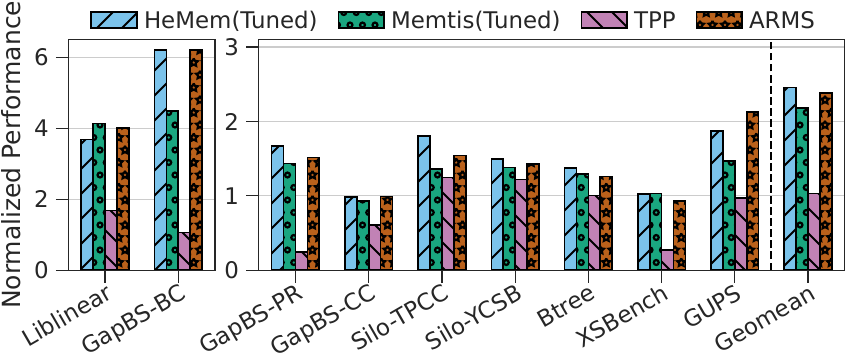}
    \caption{Comparison of tuned HeMem and Memtis, TPP and \sysname on \texttt{pmem} machine. Performance normalized to static allocation. Higher is better.}
    \label{fig:perf_comparison_scailp}
    \vspace{-1ex}
\end{figure}

\mypar{Different hardware platform.} We evaluate \sysname on \texttt{pmem} (see Table~\ref{tab:machine_spec}) which uses Optane PMem as remote memory.
Figure~\ref{fig:perf_comparison_scailp} shows \sysname' performance and other tiering systems. Once again, \sysname outperforms all existing solutions.
The performance benefit of \sysname is higher on this node since the PMem has higher latency and lower bandwidth. This experiment affirms \sysname performs well on different hardware without needing expensive parameter tuning.

\begin{figure}
    \centering
    \includegraphics[width=\linewidth]{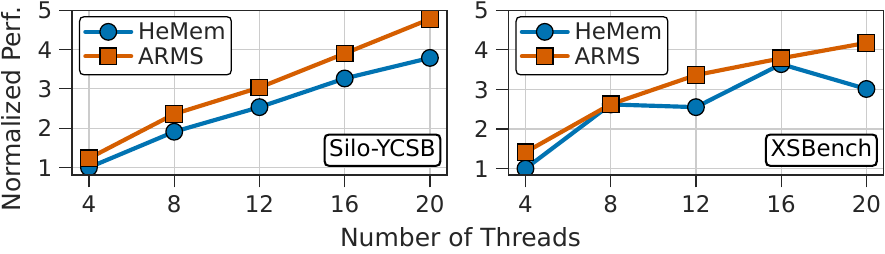}
    \caption{HeMem and \sysname performance scaling with increasing thread count for \textit{Silo} and \textit{XSBench}. Performance normalized to HeMem with 4 threads (Higher is better).}
    \label{fig:perf_scaling}
    \vspace{-3ex}
\end{figure}

\mypar{Different number of threads.} Figure~\ref{fig:perf_scaling} shows how \sysname performs for different application thread counts. \sysname achieves better performance over default HeMem consistently. The adaptive strategies in \sysname enable it to perform well with any thread count. \sysname also performs well with 20 threads: in this experiment, the sampling and policy threads share cores with the application threads.

\begin{figure}
    \centering
    \includegraphics[width=\linewidth]{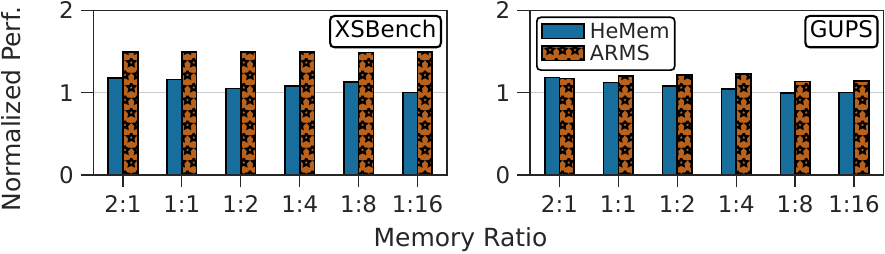}
    \caption{HeMem and \sysname performance on different memory tier ratios for \textit{XSBench} and \textit{GUPS}. Performance normalized to HeMem 1:16 memory ratio (Higher is better).}
    \label{fig:perf_memratios}
    \vspace{-1ex}
\end{figure}

\mypar{Different memory tier ratios.} We change the ratio of fast to slow tier capacities. Figure~\ref{fig:perf_memratios} shows how \sysname compares against default HeMem for different memory capacity ratios. We pick two workloads but observe similar trends for other workloads. At larger fast tier capacities, \sysname matches HeMem, but at smaller capacities (1:4 and lower), \sysname outperforms default HeMem considerably. \sysname correctly identifies the top-k hot pages for each ratio, unlike HeMem which identifies the same number of hot pages independent of the fast tier size due to a static \texttt{hot\_threshold}.

In summary, \sysname is robust and adapts to any scenario, outperforming existing systems without requiring tuning.

\section{Related work}
\label{sec:related_work}

\textbf{Page-based memory tiering systems:} Many tiering systems
are based on heuristics and use static thresholds to make  decisions. Some use recency-based policies~\cite{nimble2019,tpp2023}, some use frequency-based policies~\cite{thermostat2017,autotiering2017,hemem2021,tmts2023, memtis2023, hybridtier2025}, while others use both page access recency and frequency~\cite{tmts2023,multiclock2022,adaptivepagemigration2020,hmsdk2024}. These systems are sub-optimal as they use static thresholds and do not adapt to the workload or the underlying hardware.
The closest work is HybridTier~\cite{hybridtier2025} which uses two access counters (frequency and momentum) similar to \sysname, but still relies on static thresholds, leading to poor performance as shown in Figure~\ref{fig:perf_comparison_numa}. We observe that some parameters (e.g. 1 minute delay before demotion of second chance pages) can hurt performance.

\noindent
\textbf{Tuning tiering systems knobs:} Prior work has added adaptation to parts of a tiering system. 
Cori~\cite{cori2021} tunes the periodicity of data movement by extracting data reuse patterns from the application. Some systems determine the best migration granularity for different workloads~\cite{granularityawaremigration2018,subpagemigration2021}. Our research shows the importance of adapting all parts of a tiering system for maximum performance.
Instead of adjusting parameters, some  systems use different policies for different workloads. Heo et.al.~\cite{adaptivepagemigration2020}, propose a dynamic policy selection  that identifies the best migration policy amongst LRU, LFU and random for a given workload.
Yu et.al., build bandwidth-aware tiering systems~\cite{bwawaremigration}.

\noindent
\textbf{Other approaches.} To overcome the inaccuracies and inefficiencies in software profiling, researchers have proposed to monitor and migrate pages in the hardware~\cite{hwtiering2011,m5,neomem2024}. Others~\cite{johnnyCache2023,azure2024} have argued for hardware-managed tiering
to also support virtualized environments.
X-Mem~\cite{xmem2016} and Mira~\cite{mira2023} leverage profile-guided techniques to determine object hotness offline and make data placement decisions during allocation time. Yet, these approaches do not work well for applications whose hot set changes over time.

\section{Conclusion}
Existing tiering solutions use rigid policies along with static thresholds. As a result, these systems do not work well in all scenarios.
We identify the limitations of existing systems: inaccurate hot page classification, sluggish response to workload changes and inability to handle input noise.
We design \sysname based on robust design principles: it scores pages relatively rather than based on a threshold, smooths input to remove noise, adapts to workload access distribution changes, and estimates benefit from migrations.
We find that \sysname performs well for all workloads, on different hardware, with different inputs. Unlike prior systems, \sysname does not require tuning to achieve good performance.
\bibliographystyle{ACM-Reference-Format}
\bibliography{references}

\end{document}